# Resonant lattice Kerker effect in metasurfaces with electric and magnetic optical responses

Andrey B. Evlyukhin[1,2] and Viktoriia E. Babicheva[3]

[1]Laser Zentrum Hannover e.V., Hollerithallee 8, D-30419, Hannover, Germany
[2]ITMO University, 49 Kronverksky Ave., St. Petersburg, 197101, Russia
[3]Center for Nano-Optics, Georgia State University, Atlanta, GA, USA

**Abstract**. To achieve efficient light control at subwavelength dimensions, plasmonic and all-dielectric nanoparticles have been utilized both as a single element as well as in the arrays. Here we study 2D periodic nanoparticle arrays (metasurfaces) that support lattice resonances near the Rayleigh anomaly due to the electric dipole (ED) and magnetic dipole (MD) resonant coupling between the nanoparticles. Silicon and core-shell particles are considered. We demonstrate for the first time that, choosing of lattice periods independently in each mutual-perpendicular direction, it is possible to achieve a full overlap between the ED-lattice resonance and MD resonances of nanoparticles in certain spectral range and to realize the resonant lattice Kerker effect (resonant suppression of the scattering or reflection). At the effect conditions, the strong suppression of light reflectance in the structure is appeared due to destructive interference between electromagnetic waves scattered by ED and MD moments of every nanoparticle in the backward direction with respect to the incident light wave. Influence of the array size on the revealed reflectance and transmittance behavior is discussed. The resonant lattice Kerker effect based on the overlap of both ED and MD lattice resonances is also demonstrated.

Both plasmonic and all-dielectric nanostructures have been proposed for efficient manipulation of light at the nanoscale [1-3], with particular interest to be applied in ultra-thin functional elements, so-called metasurfaces [4-6], and in particular to control reflection from the material interfaces [7,8] and to improve photovoltaic properties [9-11]. With the typical linear dimensions of 100-200 nm, high-refractive-index (silicon) nanoparticles induce enhanced electric and magnetic moments and support Mie resonances in the visible spectral range as it has been first shown theoretically in [12] and then experimentally proved in [13] (see also [14,15]). Utilizing high-refractive-index nanoparticles gives an opportunity to obtain magnetic optical response without using metal inclusions such as split-ring resonators or core-shell particles, and avoid high non-radiative optical losses associated with metals [16]. Such all-dielectric nanostructures, for instance, nanoparticle arrays (metasurfaces), demonstrate a variety of unique effects, and in particular suppression of reflection in a pre-defined direction [17,18]. As has been shown in the earlier work [19], if electric and magnetic polarizabilities of a nanoparticle are equal each other in magnitude and phase (the first Kerker condition), light scattering from this nanoparticle is suppressed in the backward direction, and recently, it is referred as a Kerker effect. For silicon spherical nanoparticle array, ED and MD do not overlap, and only non-resonant Kerker effect is possible: antireflective properties are observed at wavelength either larger than the wavelength of the magnetic dipole (MD) resonance [12] or smaller than the wavelength of the electric dipole (ED) resonance [17,20-22]. There is a number of studies, starting with the pioneering work [23], that suggest designing nanoparticle geometry and obtaining spectral overlap of electric and magnetic dipole resonances using nanoparticles with more complex shape, such as disks [20,24-27] or cubes [28]. For the periodic arrays of resonant nanoparticles, lattice resonances (LR), appearing due to electromagnetic coupling between array nanoparticles, are realized at the wavelengths determined by the array periods in the vicinity of the Rayleigh anomalies (RA) from the side of longer wavelengths [29-41]. In the dipole approximation, the collective (lattice) resonances involve only dipole moments of the nanoparticles oriented perpendicular to the lattice wave propagation, and for the nanoparticles with ED- and MD-resonant responses, both ED-LR and MD-LR are possible. For periodic arrays of silicon nanoparticles, the LRs owing to ED and MD coupling have been firstly investigated in [12] and there non-resonant Kerker effect has been shown. Resonant properties can be utilized in sensing applications [42], narrow-band photodetectors [43], light sources [39], and solar cells [44,45].

In the present work, we propose a new concept to control resonances of the nanoparticles arranged in the periodic array, to achieve an overlap of ED-LR and MD-LR, and consequently to suppress reflectance or back-scattering from the array and metasurface as a whole. We show that by varying periods of the rectangular lattice, we can independently control positions of ED-LR and MD-LR, and the overlap of ED-LR and MD-LR causes strong suppression of reflection from the array. Our hypothesis is based on the results presented in [12] where it has been shown that for an infinite periodic array of spherical nanoparticles supported ED and MD resonant optical response the reflectance coefficient is determined by the expression

$$r = \frac{ik_d}{2S_L}\left(\alpha_{eff}^E - \alpha_{eff}^M\right), \tag{1}$$

where $S_L$ is the area of the array elementary cell, $k_d$ is the wave number in the medium around the array,

$$\alpha_{eff}^E = \frac{1}{\varepsilon_0 \varepsilon_d / \alpha^E - k_d^2 G_{xx}^0} \tag{2}$$

$$\alpha_{eff}^M = \frac{1}{1/\alpha^M - k_d^2 G_{yy}^0} \tag{3}$$

are the effective ED and MD polarizabilities of particles in the array, respectively. Here $\varepsilon_0, \varepsilon_d$ are the vacuum dielectric constant and the relative dielectric constant of surrounding medium, respectively, $\alpha^E$ and $\alpha^M$ are the ED and MD polarizabilities of single particle located in the homogenous surrounding medium, respectively, $G_{xx}^0$ and $G_{yy}^0$ are the dipole sums taking into account the influence of electromagnetic coupling between the particles in the array on their dipole moments. From (1), one can see that the reflectance is totally suppressed when the ED and MD effective polarizabilities are equal each other, and this behavior can be considered as lattice Kerker effect. The collective ED-LR and MD-LR resonances in the 2D arrays are excited if the real parts of the denominators of (2) and (3) are equal to zero [12]. If these resonances are realized at the same wavelength, we obtain the *resonant lattice Kerker effect*. Because the values of the dipole sums $G_{xx}^0$ and $G_{yy}^0$ are determined independently by the array periods in the mutual-perpendicular directions, there exists already a possibility to get the resonant Kerker effect in the periodic arrays with specially chosen periodicity. Because the collective lattice resonances are realized in the vicinity of RA [12], the resonant lattice Kerker effect ($r = 0$) is also realized in there. In our work, we consider both infinite and finite-size arrays.

**Results**

**1. Control of ED-LR and MD-LR positions**

To start with, we consider silicon spherical nanoparticles with radius *R* arranged in the rectangular lattice with periods *Dx* and *Dy* along *x*- and *y*-axis correspondently and illuminated by the plane wave with normal incidence and electric field **E** along *x*-direction (Fig. 1a). We perform both full-wave numerical modeling with finite-element method (FEM) implemented in CST Microwave Studio frequency-domain solver and the coupled-dipole equations (CDEs) approach [12]. In this work, we consider particles with only ED and MD resonances, but the concept can be extended to quadrupole resonances of the particle as well [36].

Figure 1 demonstrates spectra of the absorbance, reflectance, and transmittance calculated by CST Microwave Studio for arrays of silicon nanoparticles with different periodicity. From the analytical calculations [12] we know that for the silicon particle with $R \approx 60$ nm, ED resonance is excited around wavelength $\lambda_{ED} \approx 440$ nm and MD resonance around $\lambda_{MD} \approx 515$ nm. Furthermore, in the nanoparticle array, resonances can be excited at different wavelength according to Eqs. (2) and (3). However, for sub-diffraction periods, the influence of dipole sums is negligible, and we identify ED and MD resonances in the numerical full-wave simulations accordingly. Throughout the text, we refer to this resonances as ED and MD resonances of a single nanoparticle in the array.

By varying periods *Dx* and *Dy* independently on each other, we excite lattice resonances at the wavelength close to RA (from the side of long wavelengths): ED-LR is excited at the wavelength $\lambda_{ED-LR} \approx \lambda_{RA} = Dy$ (Fig. 1b), and MD-LR is excited at the wavelength $\lambda_{MD-LR} \approx \lambda_{RA} = Dx$ (Fig. 1c). For the array of nanoparticles in dielectric matrix with refractive index $n = 1.5$, the resonance positions are $\lambda_{ED-LR} \approx \lambda_{RA} = nDy$ (Fig. 1d) and $\lambda_{MD-LR} \approx \lambda_{RA} = nDx$ (Fig. 1g). The spectral position of the LR is close to the RA, but do not coincide with it, and more explanation about the shift and influence of lattice sum can be found for instance in [36]. See Supplementary Information, Fig. S1, for the illustration of the LR position change because of the particle effective polarizability through lattice sum with the fixed *Dy/Dx* and correspondingly fixed RA.

Importantly, Fig. 1 proves that the ED-LR and MD-LR can be tuned independently to certain wavelengths by choosing arrays periodicity in a different direction. For example, it is possible to achieve an overlap between ED-LR and MD resonance of individual particle in the array as it is shown in Fig. 1d at the wavelength $\lambda_{ED-LR} \approx \lambda_{MD} \approx 515$ nm and $Dy = 320$ nm. At these conditions, the reflectance is suppressed (Fig. 1e) and the transmittance is increased (Fig. 1f). We can consider this behavior as a *resonant lattice Kerker effect* (where the resonant effective electric dipole polarizability starts to be equal to magnetic dipole polarizability of single particles in free space). No resonant Kerker effect appears in the case of Fig. 1g-i because in this case MD-LR is shifted away from ED resonance with increasing of the period $Dx$. Additional details of this resonant lattice Kerker effect will be given in the next section.

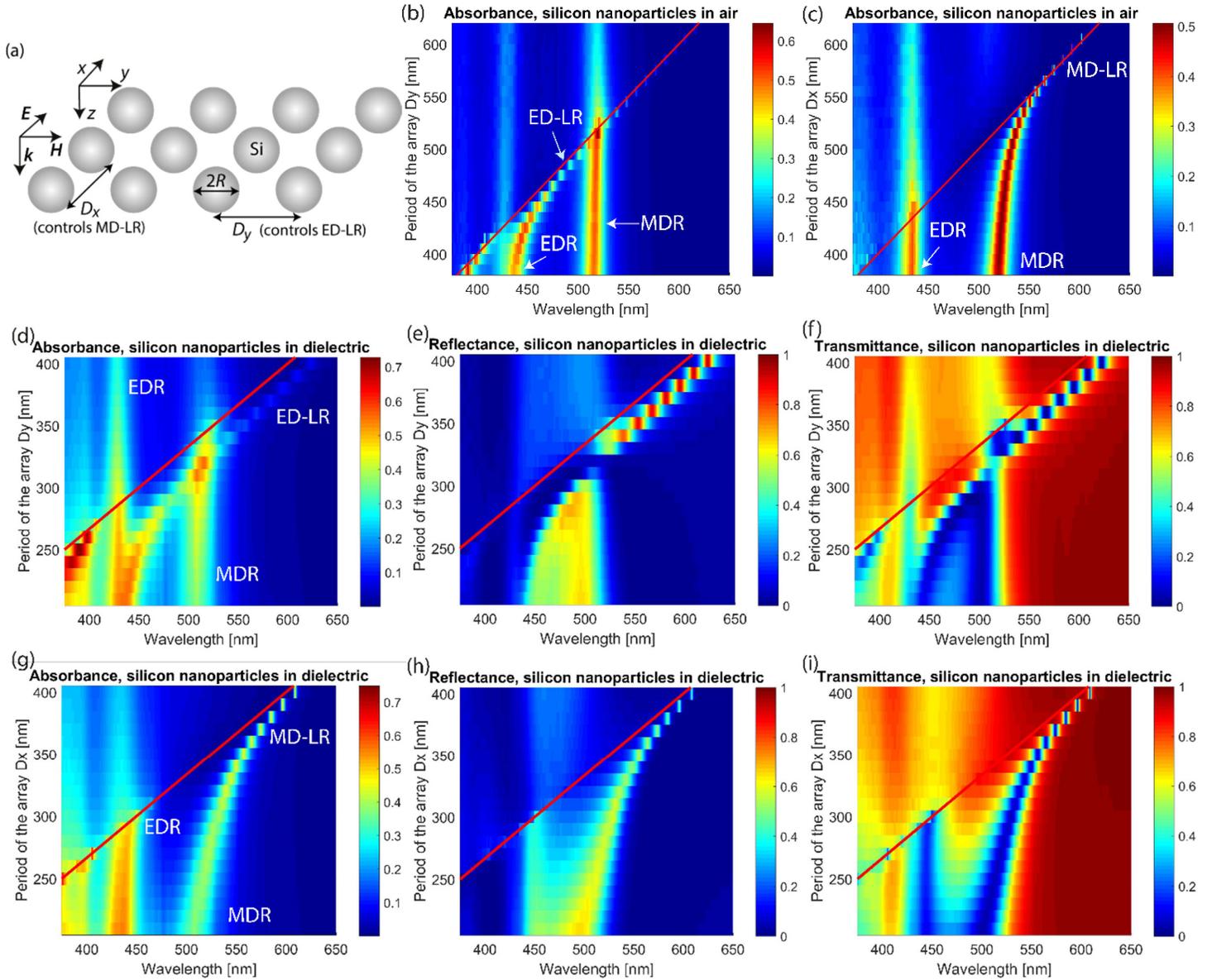

Fig. 1. (a) A periodic array of spherical silicon nanoparticles with radius $R$ in air, periods $Dx$ and $Dy$, and normal incidence of light with electric field **E** along the $x$-direction. (b) and (c) The array absorbance and the change of resonance position for different periods $Dx$ and $Dy$. EDR and MDR correspond to electric and magnetic dipole resonances respectively, and ED-LR and MD-LR to their lattice resonances counterparts. Either ED-LR or MD-LR is shifted depending on which period has been changed: ED-LR is controlled by $Dy$ and MD-LR is controlled by $Dx$. In (b), $Dx = 300$ nm and in (c), $Dy = 300$ nm. Silicon nanoparticles have $R = 60$ nm, and the arrays are in air. Red lines show positions of RA. (d) and (g) the same as (b) and (c) but for the silicon nanoparticles in a dielectric matrix with refractive index $n = 1.5$. (e) and (h) correspond to reflectance and (f) and (i) correspond to transmittance. In (d)-(f), $Dx = 210$ nm and in (g)-(i), $Dy = 210$ nm. The overlap of ED-LR and MD resonances causes an increase of absorbance, an increase of transmittance, and a decrease in reflectance.

Now we show that the excitation of the lattice resonances of ED and MD types can be controlled by the polarization of incident light. In Fig. 2a,b we demonstrate the transition between ED/MD-LRs due to changes of the linear polarization of incident light: Upon the change of the polarization, the resonances experience transition from ED-LR at φ = 0° to MD-LR at φ = 90° with the possibility of both separate or joint ED/MD-LR at the intermediate angles. The effect of polarization-independent Fano resonances in the array of core-shell nanoparticles has been shown before in [46]. However, it is not related to the lattice resonances and the transition between ED/MD-LR has not been discussed.

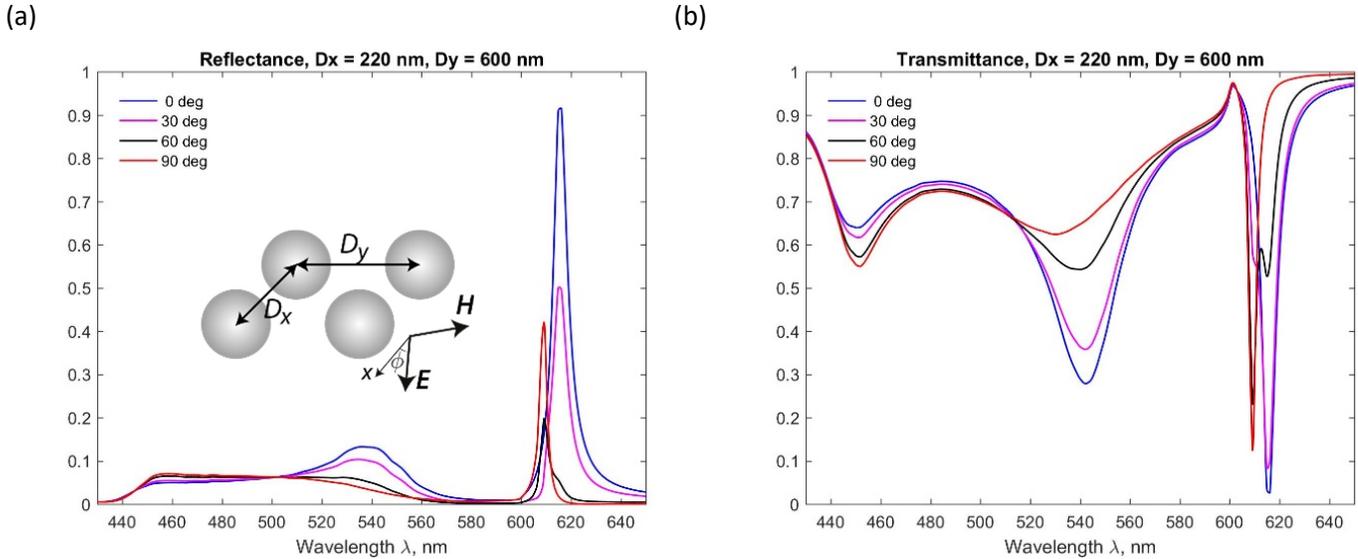

Fig. 2. (a) Reflectance and (b) transmittance that show a transition between ED-LR and MD-LR: polarization of the incident light is varied and lattice resonance at wavelength approx. 610 nm can be either ED-LR (at φ = 0°), MD-LR (at φ = 90°), or possess both separate or joint hybrid ED/MD-LR at the intermediate angles. Array periods are $D_x$ = 220 nm and $D_y$ = 600 nm, and silicon nanoparticles have R = 65 nm. Inset in (a): Schematic view of the periodic array and polarization angle φ.

## 2. Overlap of MD resonance of single particle in the array and ED-LR

Results presented in Fig. 1 have been obtained by the numerical simulation using the CST Microwave Studio without multipole decomposition, and the consideration of ED and MD resonances was only an assumption. Here we prove that in similar systems, an inclusion of only ED and MD coupling between particles sufficiently well describes their collective optical properties.

Let us consider infinite periodic arrays of silicon nanoparticles in the air with the fixed $D_x$ = 220 nm and different $D_y$. The irradiation conditions are as in Fig. 1a. Figure 3a-f presents the results calculated in the framework of two approaches including full-wave modeling FEM in the CST Microwave Studio and CDEs with only ED and MD coupling [12]. Full agreement between results in Fig. 3a-c and Fig. 3d-f confirms that only ED and MD coupling determines the main optical properties of the system. In the case $D_y$ = 450 nm (Fig. 3a,d), the features of transmission, reflection, and absorption spectra for the wavelengths > 420 nm are determined by ED, ED-LR, and MD resonances (wavelengths of 440, 485, and 530 nm, respectively) of single nanoparticles. Corresponding particle ED and MD polarizabilities are shown in Fig. 3g. For $D_y$ = 517 nm (Fig. 3b,e), the ED-LR is excited and overlaps with the MD resonance of single particles providing a strong suppression of the reflection in the system. In this case, the resonances of the effective ED and MD polarizabilities of the particle in the array coincide (Fig. 3e) that corresponds to resonant lattice Kerker effect. For larger $D_y$ = 600 nm, the resonant Kerker effect disappears (Fig. 3c,f) because the MD resonance and ED-LR are excited in the different spectral ranges (Fig. 3i).

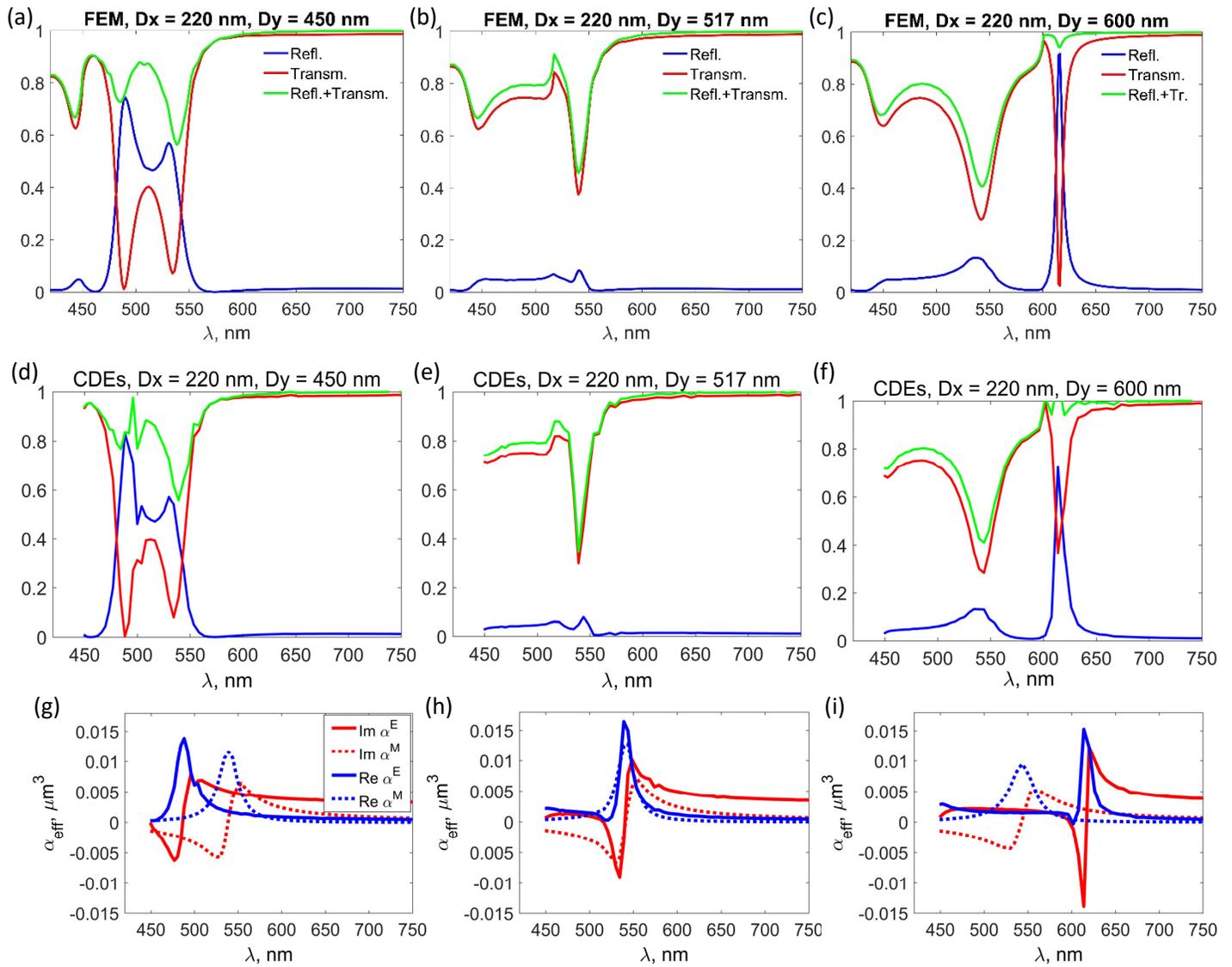

Fig. 3. Overlap of MDR and ED-LR for silicon nanoparticle with $R$ = 65 nm. (a)-(c) FEM numerical simulations of transmission and reflection coefficients (zero diffraction order). Total reflectance from the array related to Eq. (1) as $R_a = |r|^2$. (d)-(f) CDEs calculations of transmission and reflection coefficients of the array of dipoles (array is effectively infinite). Legend is the same as in (a)-(c). (g)-(i) Effective polarizabilities that correspond to the cases in (d)-(f). Legend is the same for all three panels. $Dy$ is varied and $Dx$ = 220 nm is fixed. The results show that upon spectral overlap of MDR and ED-LR for $Dy$ = 517 nm (see panel (h)), the reflectance of the array is almost zero (panels (b) and (e)). See Supplementary Information, Fig. S3, for the case of near-infrared wavelength range and near-zero optical losses in the nanoparticles.

In the calculations shown above, we have demonstrated that the overlap of ED and MD resonances causes zero reflectance in the infinite array of nanoparticles (the resonant lattice Kerker effect). Further, we show that the resonance overlap results in an increase of forward-to-backward scattering (F/B) ratio for a finite-size array of nanoparticles, and can be also regarded as resonant lattice Kerker effect. For this, we again employ both CDEs calculations and FEM numerical modeling for the finite-size array of silicon nanospheres with $R$ = 65 nm (Fig. 4). We choose $Dx$ = 220 nm and $Dy$ = 525 nm, which correspond to the overlap of ED-LR and MD resonance in the finite-size array. Coupled-dipole equations calculations (Fig. 4a-c) show that in the array of 21 x 21 nanospheres the F/B ratio can be up to 50, and the numerical modeling (Fig. 4d-f) gives the ratio up to 600 for the array of 14 x 14 nanospheres. However, even small array of 7 x 7 nanospheres provides F/B ratio above 50 (Fig. 4f). Note, that due to the resonant lattice Kerker effect these similar finite-size arrays can be used as functional elements for the efficient forward scattering of light.

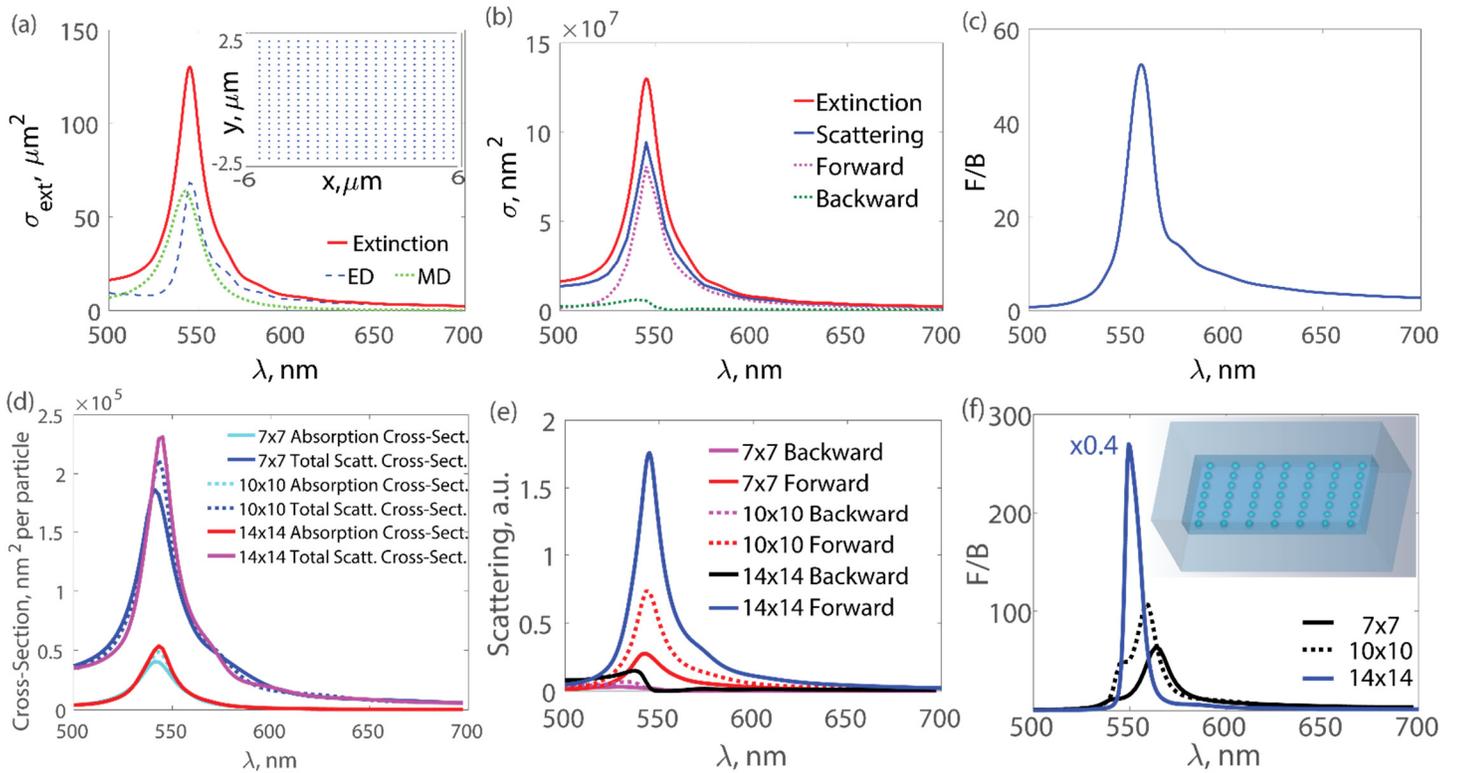

Fig. 4. Directional scattering in a finite-size array of silicon nanoparticles: $R$ = 65 nm, $Dx$ = 220 nm, and $Dy$ = 525 nm, which correspond to the overlap of ED-LR and MDR, (a)-(c) Calculations of the array of 21 x 21 nanospheres by CDEs: (a) Total extinction cross-section of nanoparticles in the array as well as its decomposition on ED and MD contributions; Insert in (a): Array of nanospheres considered in the calculations. (b) Total extinction as well as total, forward, and backward scattering cross-sections; (c) Forward-to-backward ratio that corresponds to (b). (d)-(f) Numerical modeling of the array 7 x 7, 10 x 10, and 14 x 14 nanospheres: (d) Absorption and total scattering cross-section; (e) Forward and backward scatterings; (f) Forward-to-backward scattering ratio that corresponds to (e). Insert in (f): Simulation domain with 7 x 7 nanospheres. See Supplementary Information, Fig. S4, for total power flux distributions at wavelength λ = 541 nm and 563 nm, which correspond to the peak in radar cross-section and to F/B ratio peak, respectively.

## 3. Overlap of MD-LR and ED-LR

So far, we investigated the resonant lattice Kerker effect corresponding to the overlap of the ED-LR and individual MD resonance of the particle in arrays. In this section, we demonstrate an overlap of MD-LR and ED-LR and lattice Kerker effect for two types of the arrays that consist of either silicon nanoparticles or core-shell nanoparticles. The second case is studied for the nanoparticles with spectrally overlapping MD and ED resonances in the near-infrared spectral range where the constituent materials possess small optical losses. We study this case with the purpose to illustrate that both resonances can be tuned together, and zero reflectance can be achieved. It's important to emphasize that in order to realize an overlap between MD-LR and ED-LR, it is necessary to take arrays with quadratic elementary cell and with the periods being larger than the wavelength corresponding to the MD resonance of individual particle in the array.

We perform numerical calculations of the silicon nanoparticle array with $R$ = 65 nm, $Dy$ = 600 nm, and varied $Dx$ (Fig. 5). It is expected that when $Dx$ value is close to 600 nm, the ED-LR and MD-LR almost overlap. In fact, the resonances do not overlap for $Dx$ = $Dy$ = 600 nm because of the different free space polarizabilities of the nanoparticles arrange in the lattice. A small adjustment of $Dx$ = 601.2 nm brings the resonances at the same spectral position and reflectance is significantly reduces. However, the losses prevent compensation of electric and magnetic polarizability and reflectance in not zero (the black curve in Fig. 5a). To confirm the effect of optical losses, we perform numerical modeling for the hypothetic material that has a real part of the permittivity equal to the real part of the permittivity of silicon, but its imaginary part of the permittivity is negligibly small. In this case, the reflectance is dropped down to zero (the dotted black curve in Fig. 5a).

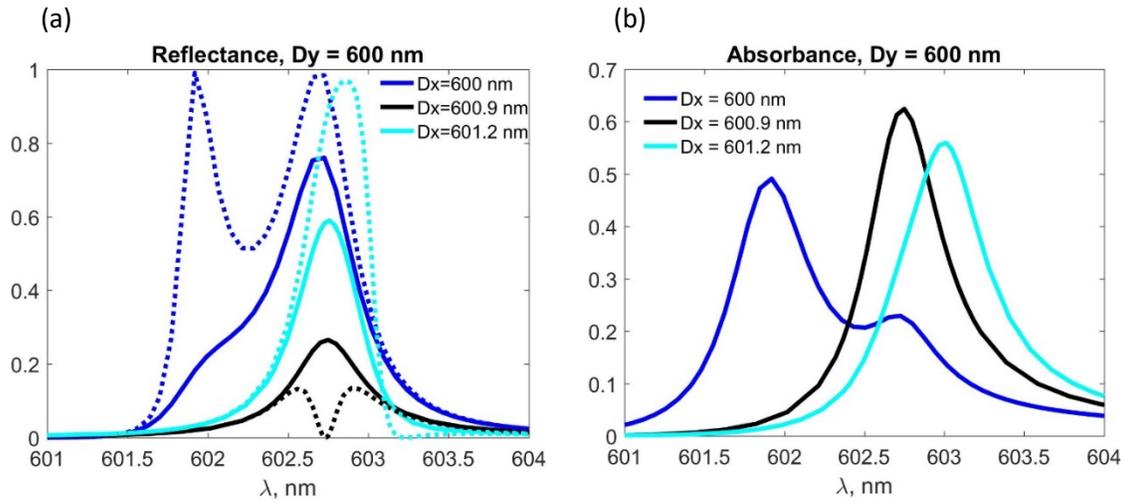

Fig. 5. Overlap of MD-LR and ED-LR in the array of silicon particles. (a) Reflectance, (b) Absorbance. $Dy$ = 600 nm and position of ED-LR is fixed. Period $Dx$ controls the position of MD-LR and brings it to the overlap with ED-LR. Solid lines correspond to the calculations for silicon particles. For the case $Dx$ = 600.9 nm, ED-LR and MD-LR are overlapped, but reflectance is not zero for silicon parameters (solid black line in (a)) and thus optical losses prevent compensation of electric and magnetic polarizability. Dotted lines correspond to the calculations with hypothetic material that has a real part of the permittivity equal to the real part of the permittivity of silicon and negligibly small optical losses. Reducing optical losses causes zero reflectance (dotted black line in (a)). Silicon nanoparticles have $R$ = 65 nm.

It is interesting to consider the case when nanoparticles have nearly overlapping ED and MD resonances, and for these purposes, we perform the analysis of the array of core-shell nanoparticles in near-infrared range (Fig. 6a,b). We study silver nanosphere with $R$ = 38 nm and dielectric shell with outer radius $R_o$ = 150 nm and refractive index $n$ = 3.5, which is close to the silicon index in the corresponding wavelength range. Such single core-shell particles possess nearly overlapping ED and MD resonances in the wavelength range 1000 – 1200 nm [46]. Here, we show that these resonances can be manipulated the same way as in dielectric nanoparticles with ED and MD resonances. In the near-infrared range, optical losses of constituent materials are relatively small, and results are expected to be similar to the case of hypothetic material studied above, overlap of resonances with nearly zero reflectance. Once the periods are equal, i.e. $Dx$ = $Dy$ = 1200 nm, ED-LR and MD-LR are spectrally close but do not overlap because of the different effective polarizabilities of the particles in the array and lattice sum contribution (Fig. 6a). The small adjustment of $Dx$ and spectral shift of MD-LR brings ED-LR and MD-LR in the overlap, which results in the zero-reflectance band at 1215 – 1222 nm.

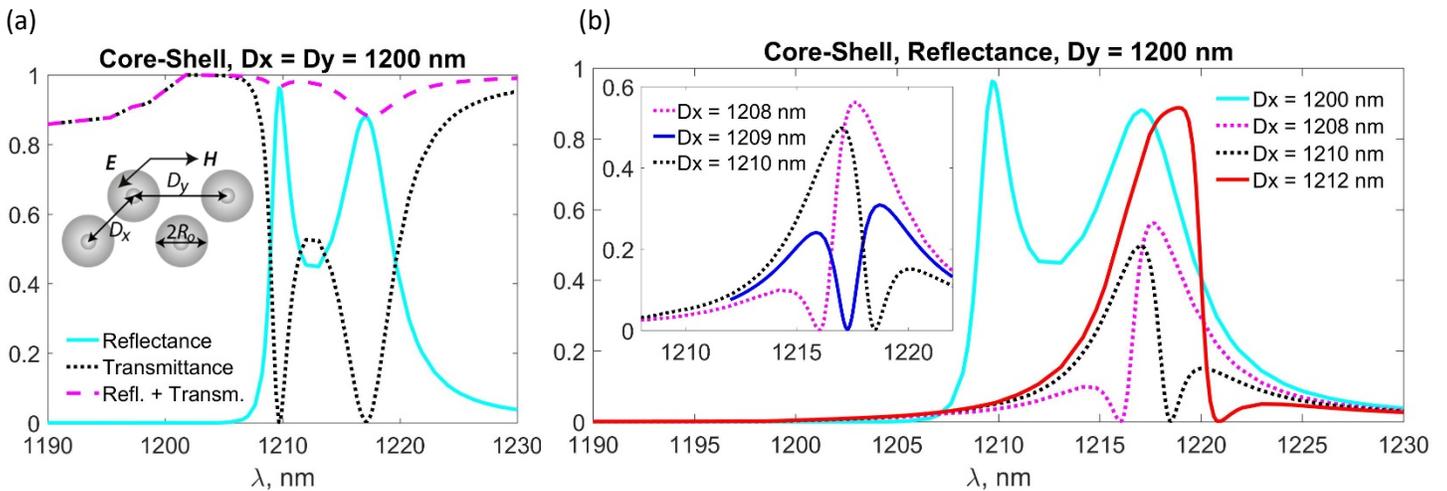

Fig. 6. Overlap of ED-LR and MD-LR for the array of core-shell nanoparticles. (a) Reflectance, transmittance, and absorbance for $Dx$ = $Dy$ = 1200 nm. Inset in (a): Schematic view of core-shell nanoparticle array. See Supplementary Information, Fig. S5, for $Dx$ = $Dy$ = 800 nm and $Dx$ = $Dy$ = 1050 nm. (b) Reflectance for $Dy$ = 1200 nm. Inset in (b): Comparison that includes $Dx$ = 1209 nm. See Supplementary Information, Fig. S6, for logarithmic scale and spectra with other parameters of $Dy$.

**Conclusion**

Lattice resonance is a resonant excitation of the particle dipoles arranged in the periodic array, and we have shown that spectral overlap between ED-LR and MD dipole of single nanoparticles in arrays; and also between ED-LR and MD-LR causes a directional scattering and the Kerker effect similar to that one upon an overlap of ED resonance and MD resonance of the single particle. We considered only ED and MD resonances, but the concept can be extended to quadrupole resonances as well. First, we have shown that varying periods of the rectangular lattice of silicon spherical nanoparticles, we can independently control positions of ED-LR and MD-LR, and we have demonstrated it for both free-standing particles and particles in the dielectric matrix, which shifts both RAs and LRs in comparison to the free-standing case. Second, we have shown that the overlap of ED-LR and MD resonance of single particles causes near-zero reflection from the array, and we have introduced the resonant lattice Kerker effect linking it to known Kerker effect for single particles. We have demonstrated this lattice Kerker effect for array with a finite number of nanoparticles (7 to 21 in one direction) and have shown that the ratio of forward to backward scattering from such finite-size array can be up to 600, and patches of such arrays can be used for efficient light harvesting or to improve performance of ultra-thin elements based on metasurfaces. We have also demonstrated a decrease of reflection upon overlap of ED-LR and MD-LR with the reflectance down to zero in the case of negligible non-radiative losses in the nanoparticles, which can take place at the near-infrared range. We have also confirmed this effect for nanoparticles made of a metal core and dielectric shell (core-shell nanoparticles) designed to have an overlapping ED and MD resonances in the considered spectral range. We would like to note that with the nanoparticles of more complex shapes like disks and cones, it is possible to excite ED resonance at larger wavelength than MD resonance, which in principle allows for overlapping ED and MD-LR and achieve resonant lattice Kerker effect in the array of such particles.

**Acknowledgment**. The authors acknowledge financial support from the Deutsche Forschungsgemeinschaft (Germany), the project EV 220/2-1 and from the Russian Science Foundation (Russian Federation), the project 16-12-10287.

# Supplementary Information for
## Resonant lattice Kerker effect in metasurfaces with electric and magnetic optical responses

A.B. Evlyukhin and V.E. Babicheva

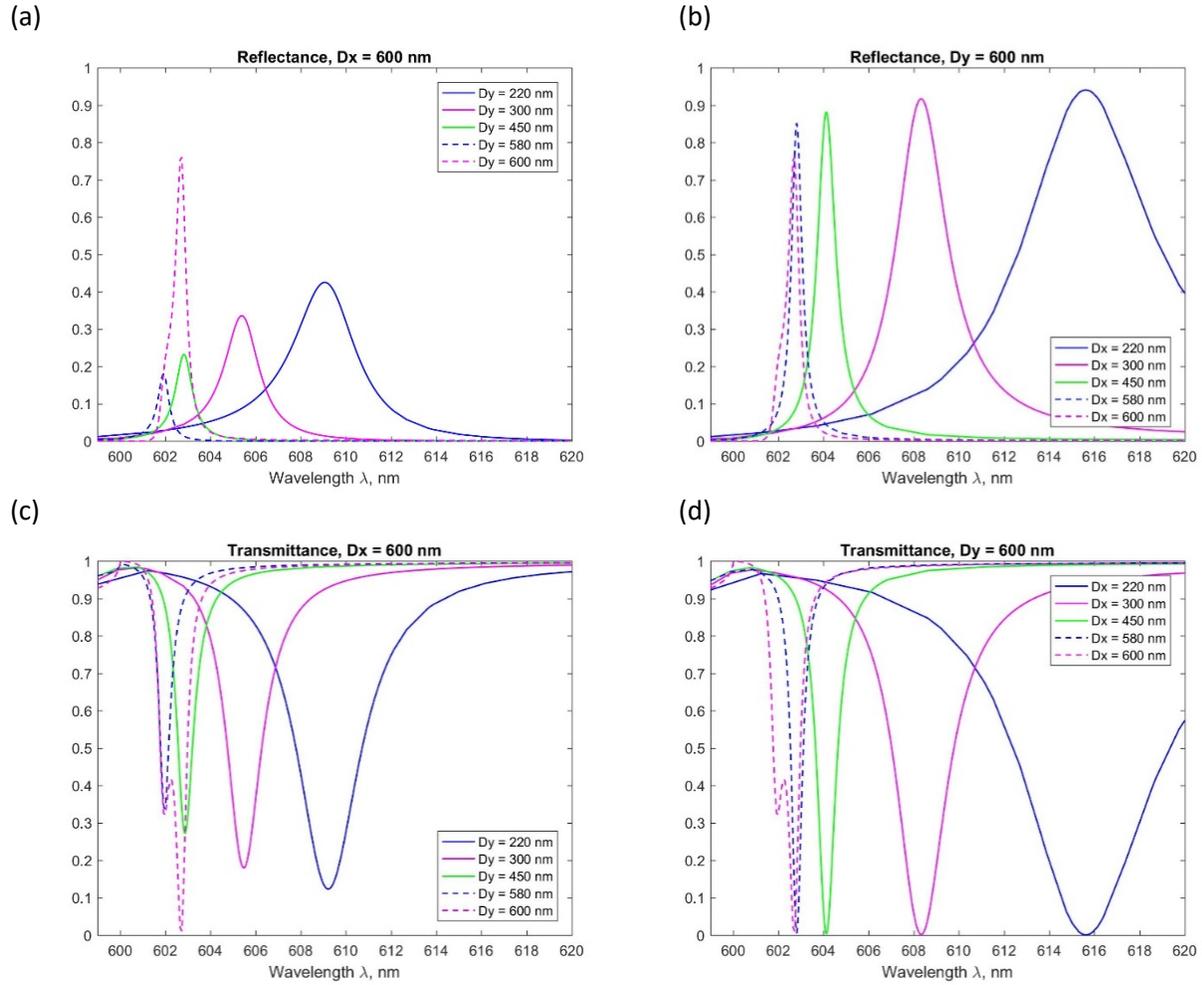

Fig. S1. Control of ED-LR or MD-LR position by the change of particle effective polarizability through lattice sum with the fixed Dy/Dx and correspondently fixed RA. (a) Reflectance of the array with MD-LR (Dx = 600 nm) at the different Dy: the resonance is red-shifted for the dense array (Dy = 220 nm) and close to the $\lambda_{RA}$ = Dx = 600 nm for the sparse array with Dy = 600 nm. (b) is the same as (a) but for Dy = 600 nm and ED-LR. Silicon nanoparticles have R = 65 nm.

(a)

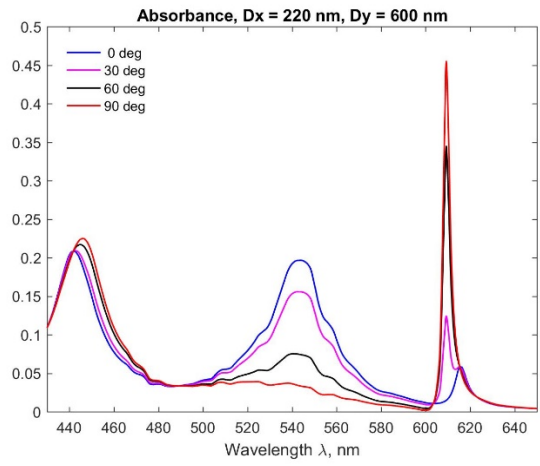

(b)

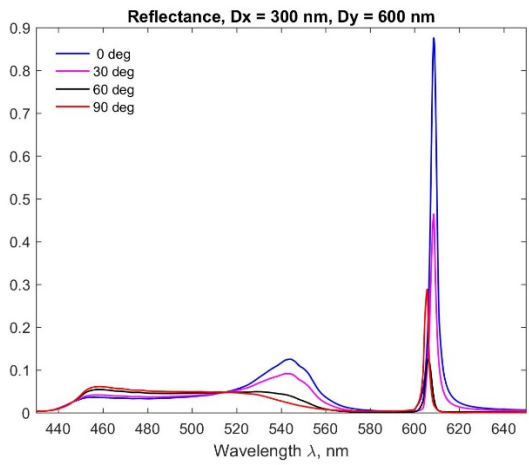

(e)

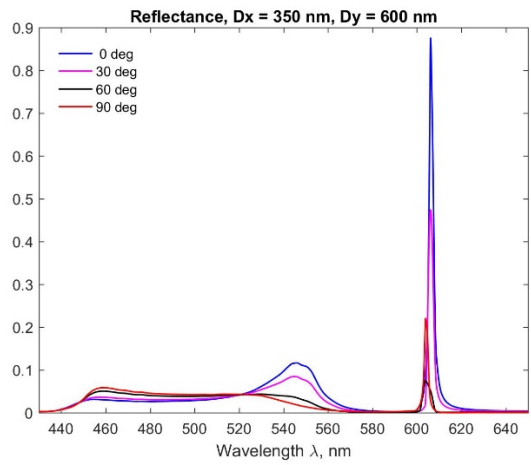

(c)

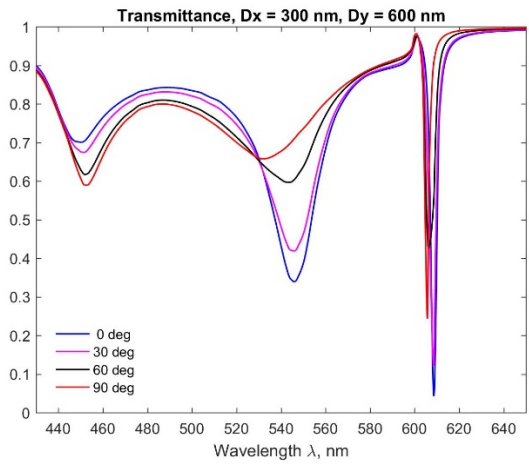

(f)

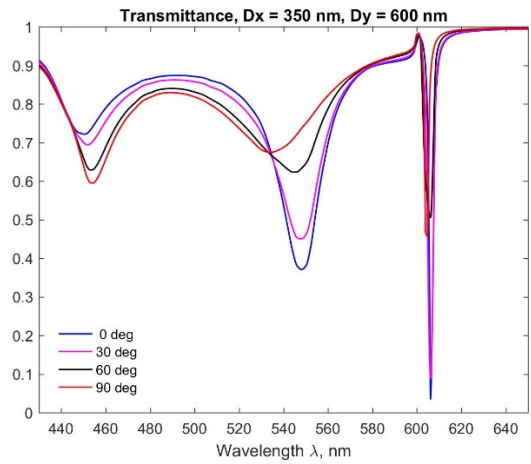

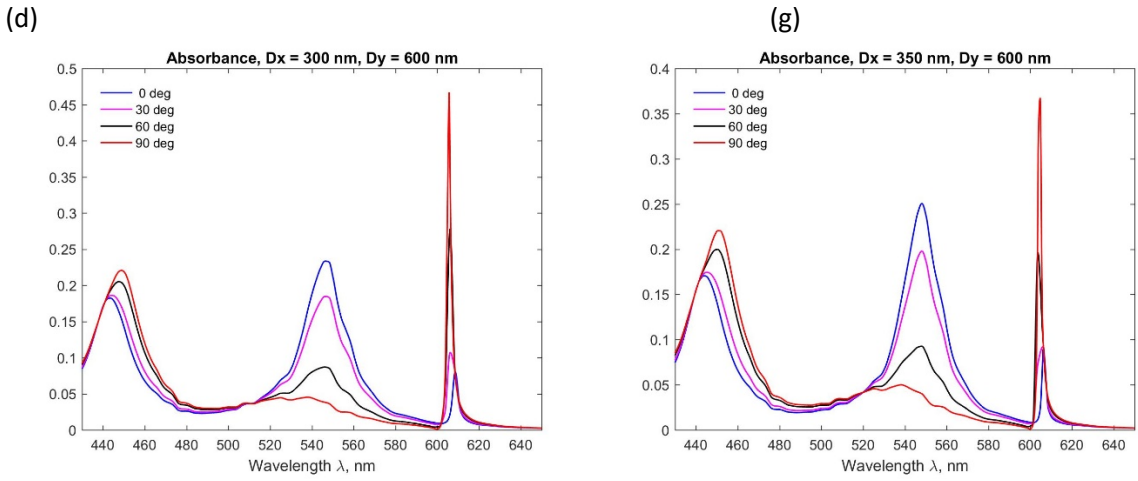

Fig. S2. Transitions between ED-LR and MD-LR: polarization of the incident light is rotated and lattice resonance at wavelength approx. 610 nm can be either ED-LR (at φ = 0°), MD-LR (at φ = 90°), or possess both separate or joint hybrid ED/MD-LR at the intermediate angles. (a) Absorbance for Dx = 220 nm, Dy = 600 nm; (b) Reflectance, (c) Transmittance, and (d) Absorbance for Dx = 300 nm, Dy = 600 nm; (e) Reflectance, (f) Transmittance, and (g) Absorbance for Dx = 350 nm, Dy = 600 nm.

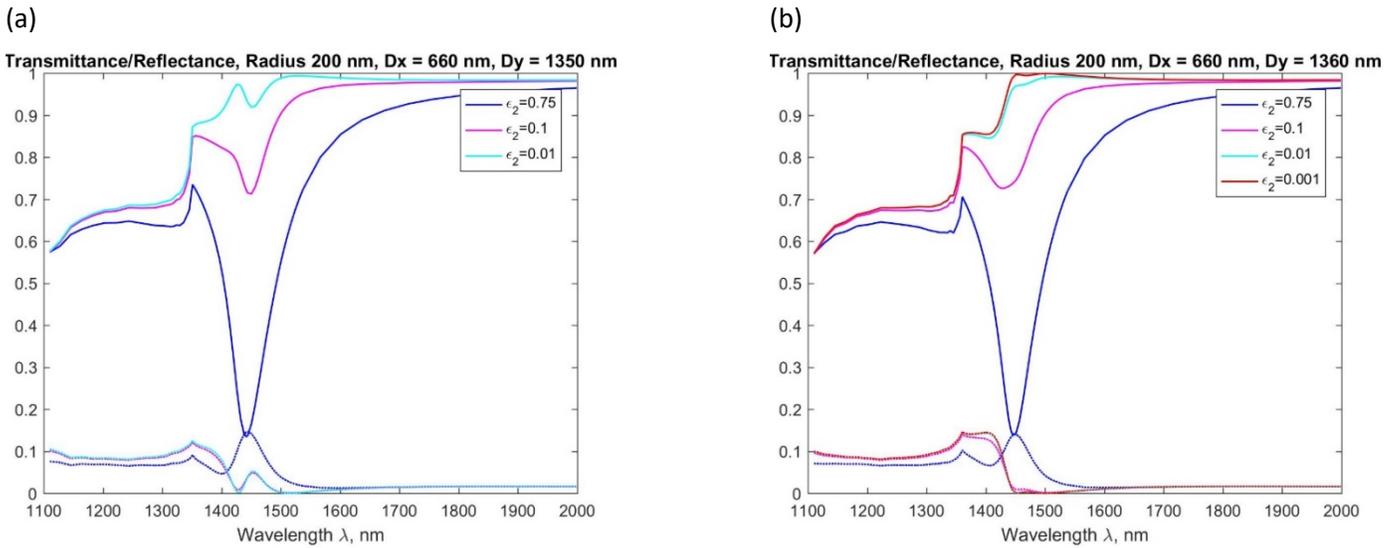

Fig. S3. Transmitance (solid lines) and reflectance (dotted lines) spactra for the array with (a) Dy = 1350 nm and (b) Dy = 1360 nm. Calculations are perform for the hypothetical material with the real part of permitivity $\varepsilon_1$ equal to $\varepsilon_1$ of silicon, and imaginary part $\varepsilon_2$ is varied. Particle radius R = 200 nm and Dx = 660 nm. For the case of small optical losses ($\varepsilon_2 < 0.1$), refleranct is suppred down to zero.

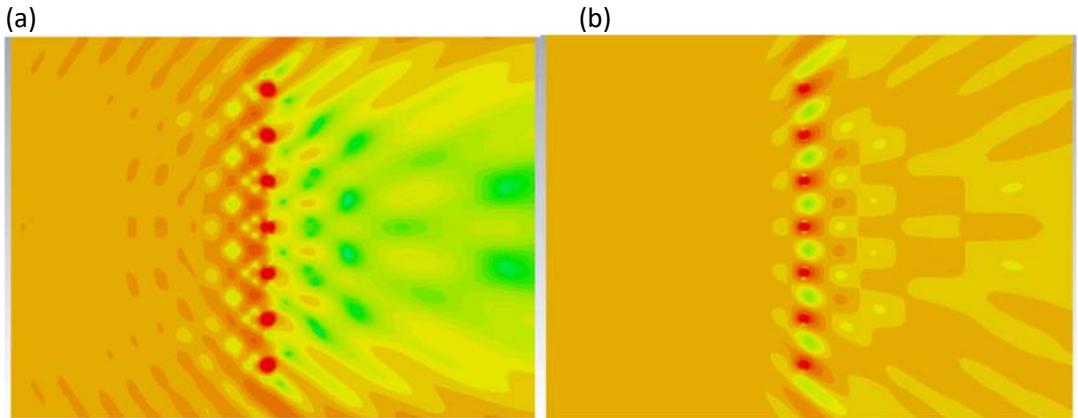

Fig. S4. (a) Total power distribution at the wavelength λ = 541 nm, which corresponds to the peak in radar cross-section, (b) Total power distribution at the wavelength λ = 563 nm, which corresponds to F/B ratio peak. The array is 7x7 nanospheres with R = 65 nm, Dx = 220 nm, Dy = 525 nm.

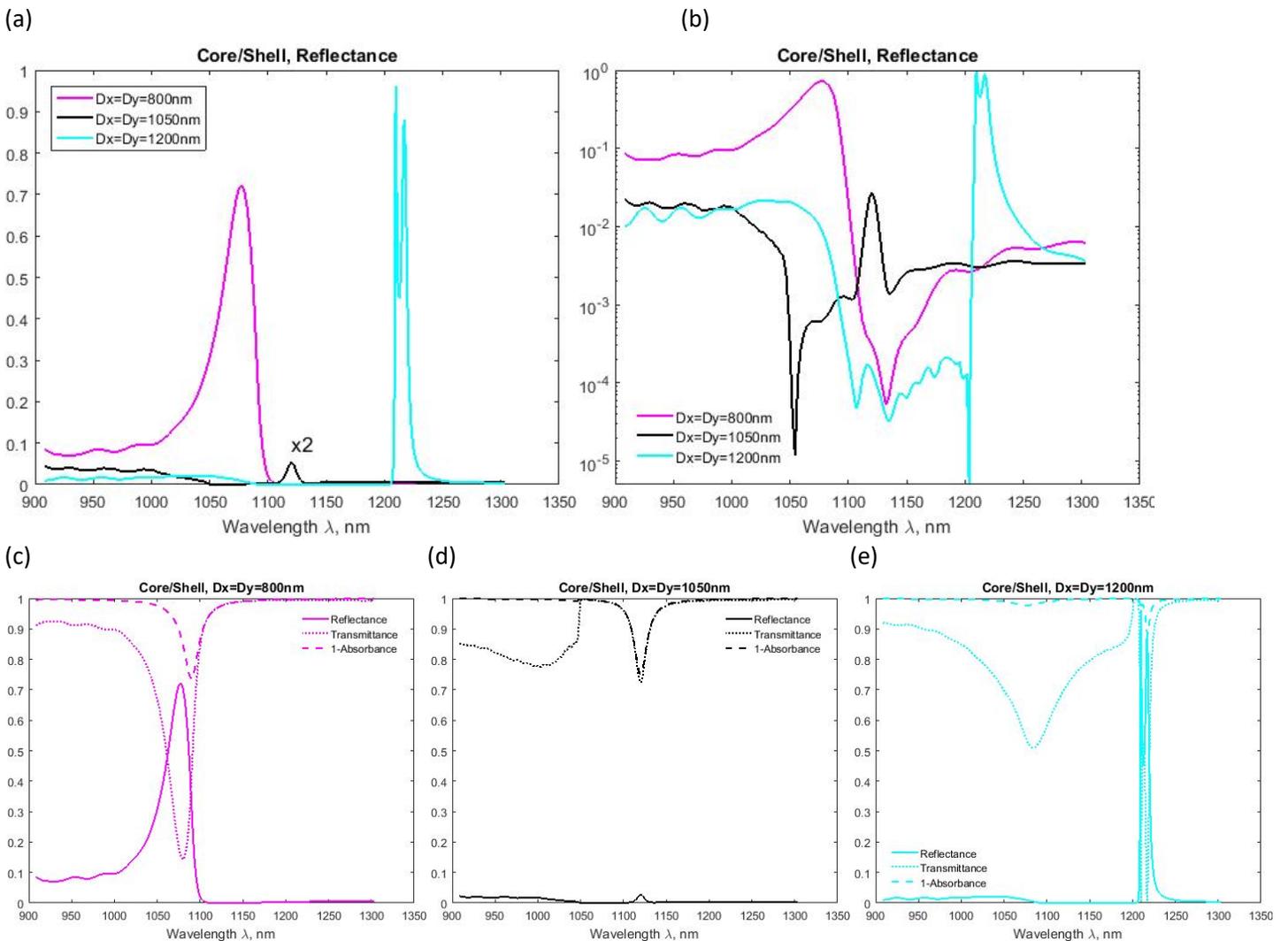

Fig. S5. Overlap of ED-LR and MD-LR for the array of core-shell nanoparticles. (a) and (b) Reflectance spectra in liner and logarithmic scales, respectively. (c)-(e) Spectra of reflectance, transmittance, and absorbance for (c) Dx = Dy = 800 nm, (d) Dx = Dy = 1050 nm, and (e) Dx = Dy = 1200 nm. The difference between R+T and 1-Abs is associated with the fist diffraction order, which is significant for wavelength λ < max(Dx, Dy).

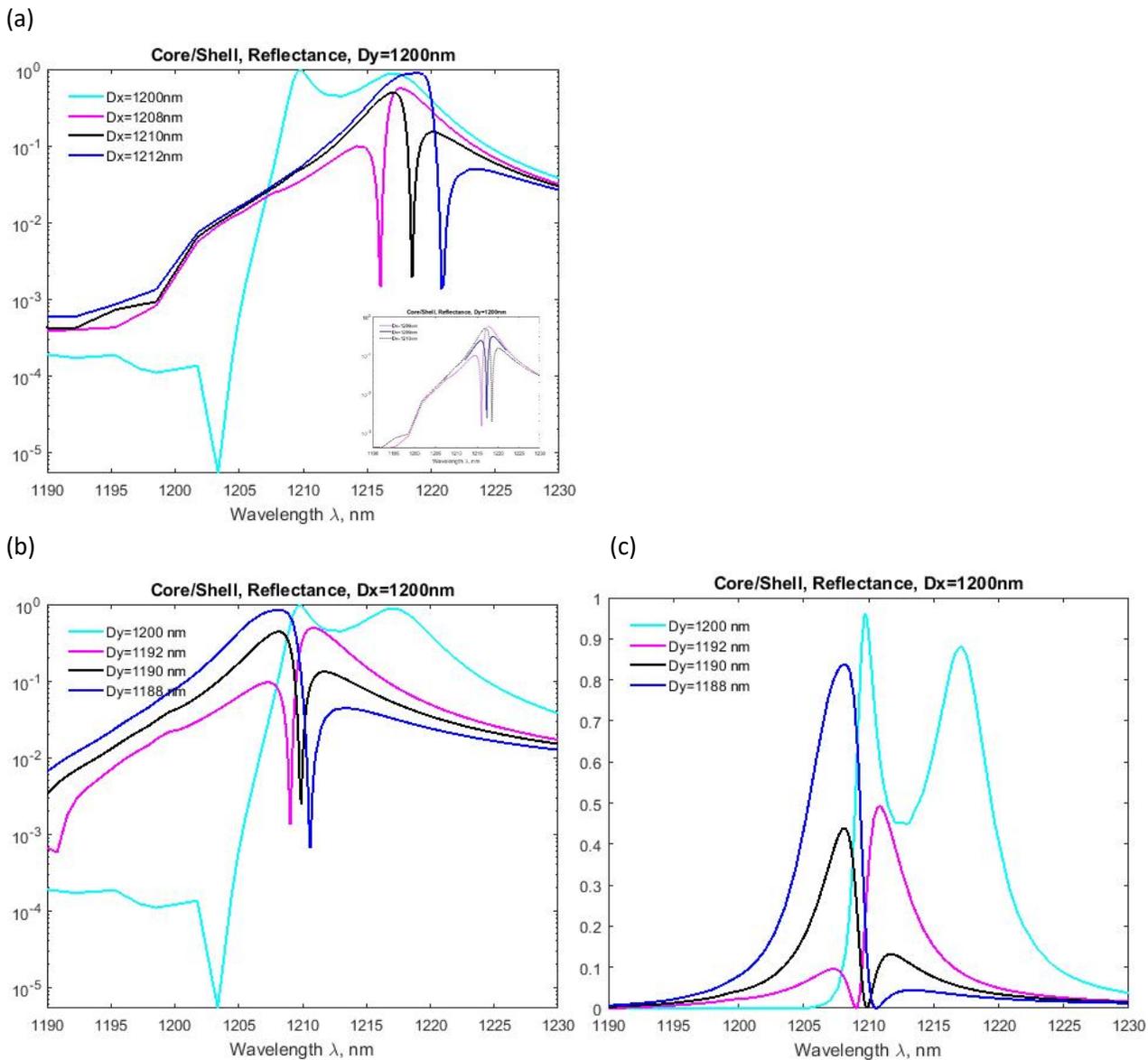

Fig. S6. Overlap of ED-LR and MD-LR for the array of core-shell nanoparticles. (a) Logarithmic scale for Dy = 1200 nm, (b) Logarithmic and (c) Linear scale for Dx = 1200 nm.